\documentclass[aps,amsmath,prd,showpacs,nofootinbib,amssymb,twocolumn]{revtex4}

\usepackage{amssymb,graphicx,color,dcolumn,bm,slashed,tabularx,ulem}
\makeatletter

\def\roughly#1{\mathrel{\raise.3ex\hbox
{$#1$\kern-.75em\lower1ex\hbox{$\sim$}}}}

\begin{document}

\title{Thermal transport of the solar captured dark matter and its impact on the indirect dark matter search}

\author{Chian-Shu~Chen$^{1,3}$\footnote{chianshu@gmail.com}, Guey-Lin Lin$^{2}$\footnote{glin@cc.nctu.edu.tw}, and Yen-Hsun Lin$^{2}$\footnote{chris.py99g@g2.nctu.edu.tw}}
  \affiliation{$^{1}$Institute of Physics, Academia Sinica, Nangang, Taipei, Taiwan 11529\\
$^{2}$Institute of Physics, National Chiao Tung University, Hsinchu 30010, Taiwan\\
$^{3}$Physics Division, National Center for Theoretical Sciences, Hsinchu 30010, Taiwan}

\date{Draft \today}

\begin{abstract}
We study the thermal transport occurring in the system of solar captured dark matter (DM) and explore its impact on the DM indirect search signal.
We particularly focus on  the scenario of self-interacting DM (SIDM). The flows of energies in and out of the system are caused by solar captures via DM-nucleon and DM-DM scatterings, the energy dissipation via DM annihilation, and the heat exchange between DM and solar nuclei. 
We examine the DM temperature evolution and demonstrate that the DM temperature can be higher than the core temperature of the Sun if the DM-nucleon cross section is sufficiently small
such that the energy flow due to DM self-interaction becomes relatively important.
We argue that the correct DM temperature should be used for accurately predicting the DM annihilation rate, which is relevant to the DM indirect detection.  

\end{abstract}

\maketitle
Dark matter (DM) composes about 25\% of the energy density of in the universe and plays an important role in the structure formation. 
It was shown that if the galactic halo is constituted by weakly interacting massive particles (WIMPs), there is a high possibility that these WIMPs are captured by the Sun~\cite{Silk:1985ax,Srednicki:1986vj,Spergel:1984re,
Press:1985ug,Griest:1986yu,Gould:1987ju,Gould:1991hx,
Zentner:2009is,Chen:2014oaa}. In general there are thermal energy flows between the captured DMs and the nuclei in the Sun. Microscopically, such flows are caused by particle scatterings. 
For collisionless cold dark matter (CCDM), the scatterings are only between the DM and solar nuclei.  Regarding the huge difference in abundance between the two, 
it is reasonable to take the DM temperature to be identical to the core temperature of the Sun. However, the scenario of SIDM can change the picture dramatically. In such a scenario, the energy transports via DM self-capture and DM-nucleus scattering compete with each other. In particular, if DM-nucleus interaction is much weaker than expected, the thermal exchange between DMs and nuclei would be much less efficient. As a result, the DM temperature can be distinct from the solar core temperature. 

It is worth mentioning that the DM abundance is not much affected with a suppressed DM-nucleon cross section $\sigma_{\chi p}$ provided SIDM is considered~\cite{Chen:2015bwa}. The accumulated DM abundance is determined by the balance among the capture, the evaporation, and the annihilation rates. Earlier works on such processes only consider DM-nucleon interactions. The new effects from DM self-interaction are investigated recently~\cite{Zentner:2009is,Albuquerque:2013xna,Chen:2014oaa,Chen:2015bwa}. The consideration of SIDM comes from the discrepancies between the numerical N-body simulations using the hypothesis of CCDM and the astrophysical observations on the small structure of the universe~\cite{Spergel:1999mh}. The CCDM simulations~\cite{Navarro:1996gj} predict cuspy profiles in the center regions of galaxies, which conflict with flatten cores found in our Milky Way (MW)~\cite{Walker:2011zu}, other nearby dwarfs~\cite{Oh:2010ea}, and low luminous galaxies~\cite{Gentile:2004tb,KuziodeNaray:2007qi}. There are additional puzzles concerning the sizes of subhalos. The observed MW satellites are hosted by much less massive subhalos compared to sizes of the most massive subhalos arising from simulations. The absence of such massive DM subhalos is referred to as the ``too big to fail'' problem in the galaxy formation. The existence of SIDM (in particular those involving the characteristic velocity-dependence cross sections~\cite{Feng:2009hw,Buckley:2009in,Tulin:2012wi}) is one of the solutions to alleviate these inconsistencies. 

\begin{figure*}[t]
	\begin{centering}
		\includegraphics[width=0.33\textwidth]{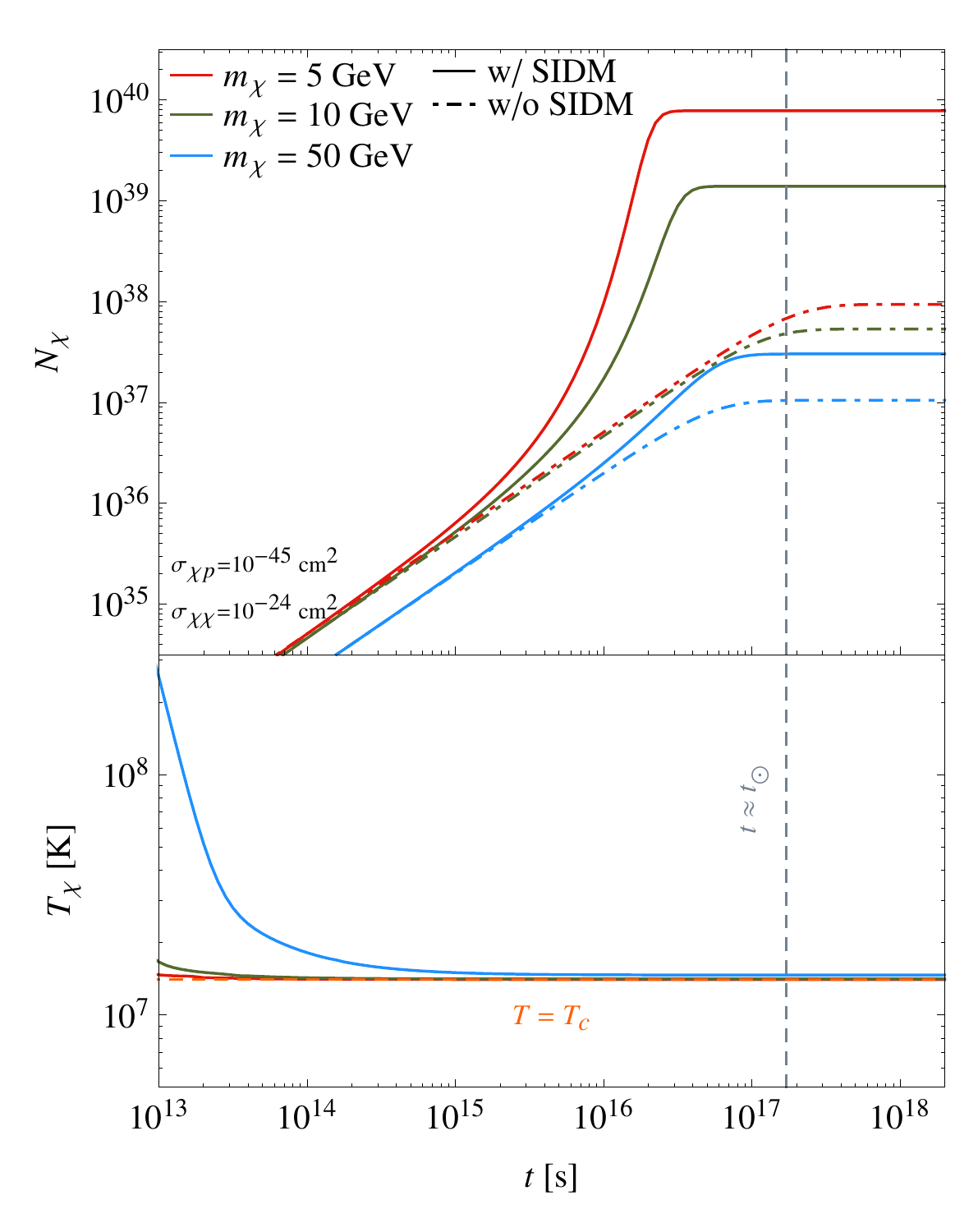}\quad\includegraphics[width=0.33\textwidth]{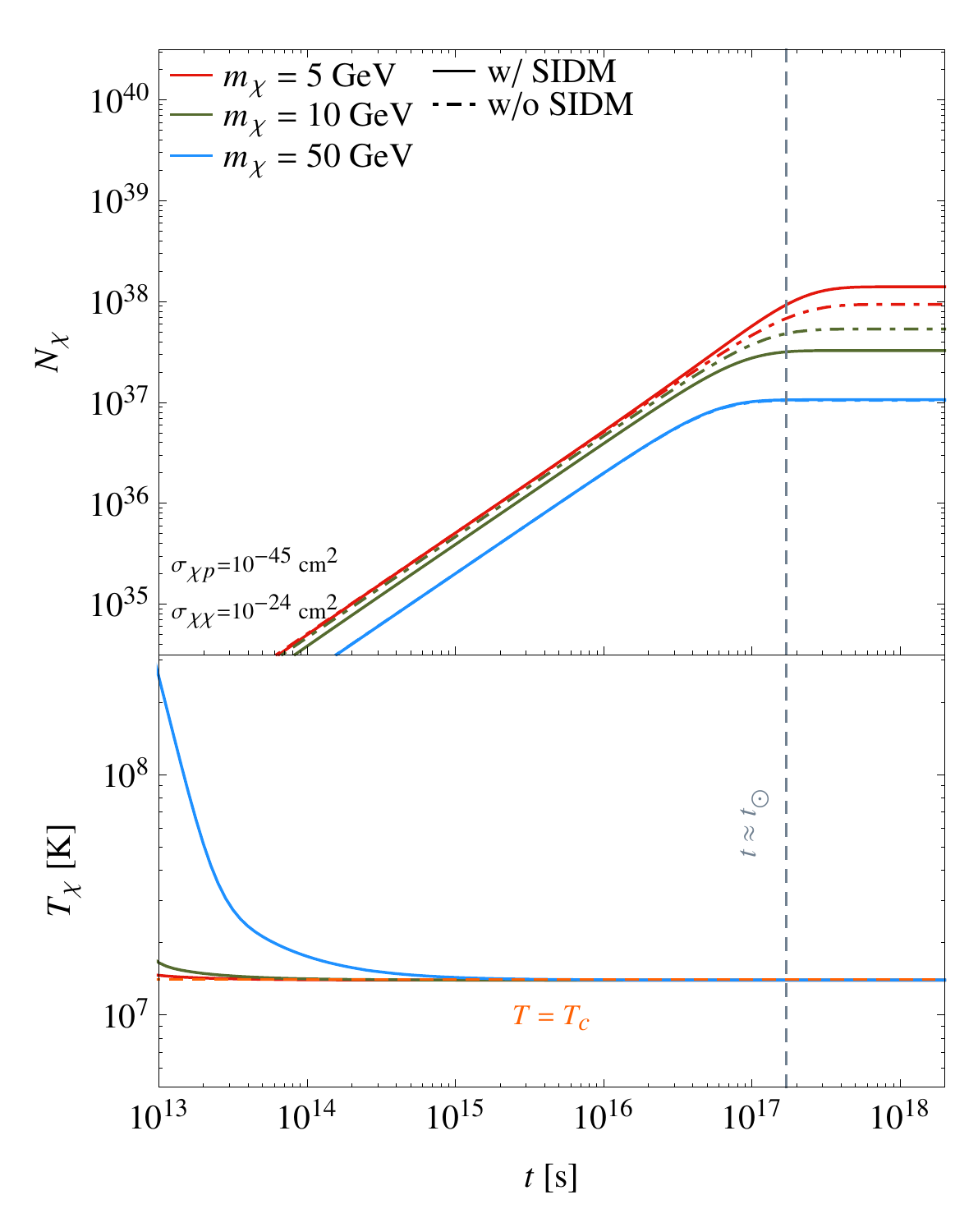}
		\includegraphics[width=0.33\textwidth]{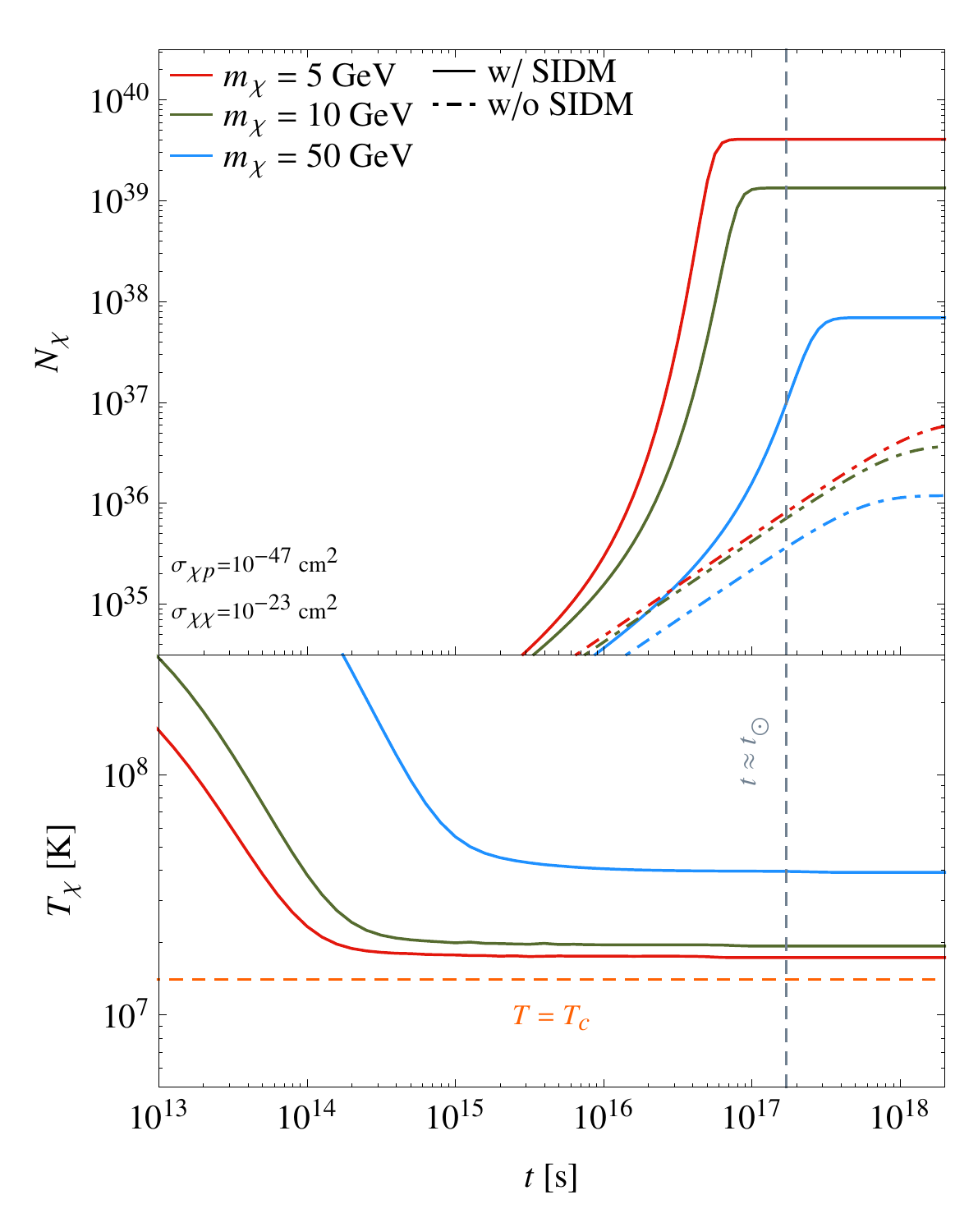}\quad\includegraphics[width=0.33\textwidth]{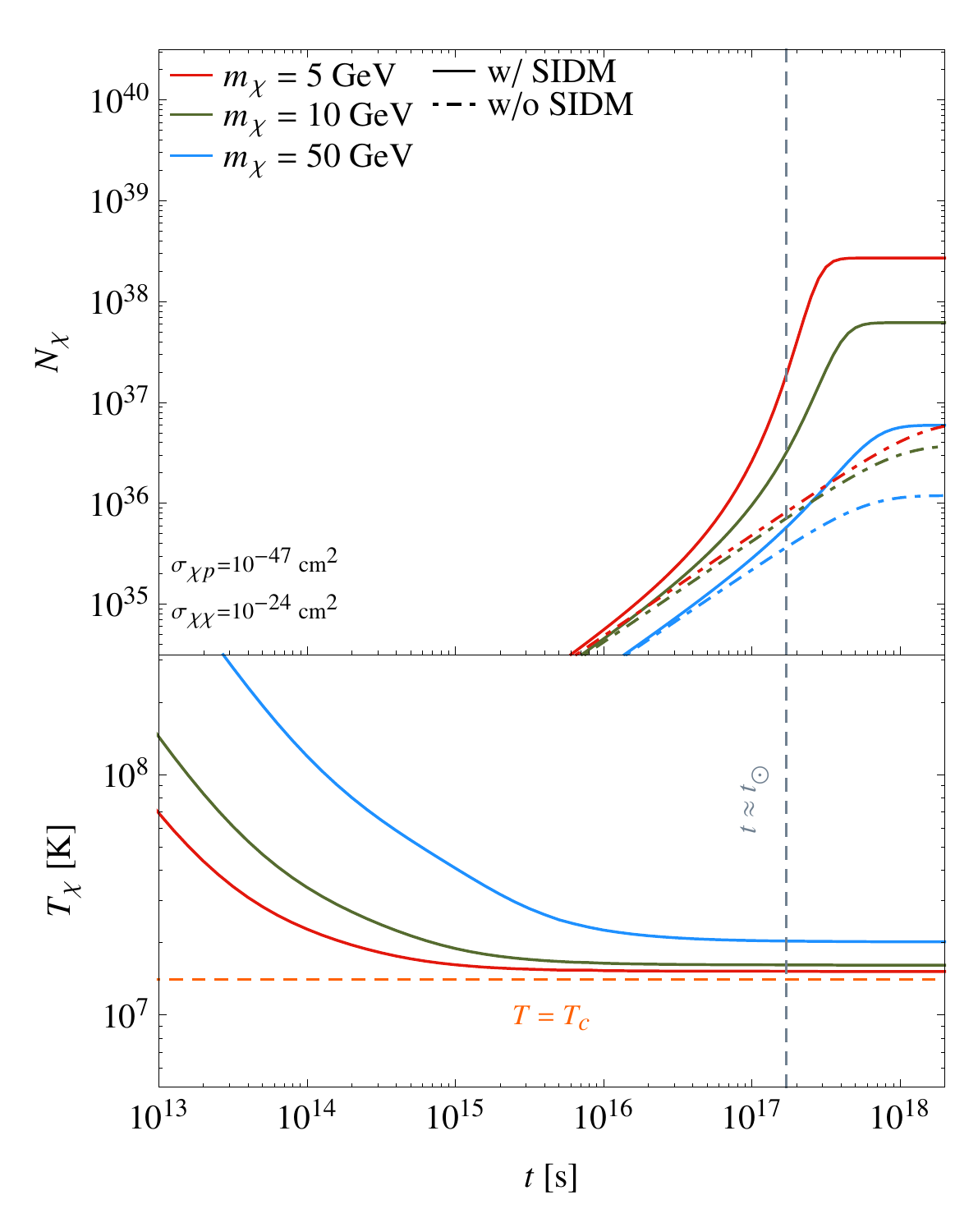}
		\par\end{centering}
	\protect\caption{The time evolutions of $N_{\chi}$ and $T_{\chi}$ for $\sigma_{\chi p} = 10^{-45}~{\rm cm^2}$ (upper pannel) and $\sigma_{\chi p} = 10^{-47}~{\rm cm^2}$ (lower panel) with $\sigma_{\chi\chi} = 10^{-23}~{\rm cm^2}$ (left) and $10^{-24}~{\rm cm^2}$ (right).\label{temperature}}
\end{figure*}

In this paper, we investigate the thermal energy transport between the trapped DM and the nuclei in the Sun, as well as the energy flows due to DM captures. 
With DM self-interactions taken into account, the DM number trapped in the Sun evolves according to 
\begin{eqnarray}
\frac{dN_{\chi}}{dt} = C_{c}+C_{s}N_{\chi} - C_{a}N_{\chi}^{2}~,\label{eq:N_evol}
\end{eqnarray}
while the energy flows in and out of the system of trapped DM in the Sun are governed by
\begin{eqnarray}
\frac{d\left(N_{\chi}E_{\chi}(t)\right)}{dt} & =J_{c}+(J_\chi+J_{s})N_{\chi} - J_{a}N_{\chi}^{2}~.\label{thermaltransport}
\end{eqnarray}
The coefficients $C_{c,s}$ are referred to as the DM capture rates due to DM-nucleon scattering cross section $\sigma_{\chi p}$ and DM-DM scattering cross section $\sigma_{\chi\chi}$, respectively. The coefficient $C_{a}$ is related to DM annihilation rate in the Sun. In this work, we shall focus on the self-interaction dominant scenario with 
$C_s^2\gg 4C_cC_a$~\cite{Zentner:2009is}. The left hand side of Eq.~(\ref{thermaltransport})
is the total kinetic energy of the trapped DM with $E_{\chi}(t)$ the average kinetic energy of an individual DM. The factors  $J_{c,s,a}$ are thermal transport coefficients corresponding to
coefficients $C_c$, $C_s$, and $C_a$, respectively. The coefficient $J_{\chi}$ describes the heat exchange between trapped DM and the nuclei in the Sun.
We do not concern the mass range where DM evaporation rate is significant.  In this case, the DM abundance in the Sun is severely reduced  
so that the signal strength of DM annihilation is suppressed. 

Since we aim at studying the temperature evolution of the trapped DM, it is important to compare mean collision time between a pair of trapped DMs and that between a trapped DM and nucleus in the Sun.
We make the comparison for
the cases of spin-independent (SI) and spin-dependent (SD) DM-nuclei scattering cross sections, respectively.   
We first note that the mean collision time between two DMs in the Sun is
\begin{eqnarray}
\tau_{\chi\chi}(t) \simeq \frac{V_{\odot}}{N_{\chi}(t)\sigma_{\chi\chi}\bar{v}},~
\label{collision:DM_DM}
\end{eqnarray} 
while the mean collision time between DM and nucleus in the Sun is 
\begin{eqnarray}
\tau_{\chi\odot} \simeq\frac{V_{\odot}}{\sum_i N_i\sigma^{\rm SI}_{\chi A_i}\bar{v}}~
\label{collision:DM_nuclei}
\end{eqnarray} 
for the case of spin-independent cross section with $V_{\odot}$ the solar volume, $\bar{v}$  the average velocity of trapped DMs, $N_i$ the number of nucleus $i$ in the Sun, and $\sigma^{\rm SI}_{\chi A_i}$ the spin-independent DM-nucleus scattering cross section. We neglect possible numerical factors on the right hand side of Eqs.~(\ref{collision:DM_DM}) and (\ref{collision:DM_nuclei}) for this order of magnitude estimation.   
The time scale $\tau^{\rm eq}_{\chi}$ for DMs in the Sun to reach thermal equilibrium can be estimated 
by the condition $\tau^{\rm eq}_{\chi}\simeq \tau_{\chi\chi}(\tau^{\rm eq}_{\chi})$. 
To solve this approximated equation, we first assume that $N_{\chi}(\tau^{\rm eq}_{\chi})$ is still far from the maximal value of $N_{\chi}$ and verify 
this assumption later. In this case, one can show that $N_{\chi}(\tau^{\rm eq}_{\chi})=C_c \tau^{\rm eq}_{\chi}$ for $C_s^2\gg 4C_cC_a$. With this input, we obtain 
$\tau^{\rm eq}_{\chi}=\sqrt{V_{\odot}/C_c\sigma_{\chi\chi}\bar{v}}$.

For further discussions, we shall take $m_{\chi}=10$ GeV as a benchmark. For such a DM mass, we have $\sum_i N_i\sigma^{\rm SI}_{\chi A_i}\simeq 40 N_H\sigma^{\rm SI}_{\chi p}$ by assuming isospin invariant DM-nucleon couplings. 
Hence we may write $r\equiv \tau^{\rm eq}_{\chi}/ \tau_{\chi\odot}=40 N_H\sigma^{\rm SI}_{\chi p}/\sigma_{\chi\chi}N_{\chi}(\tau^{\rm eq}_{\chi})$. To simplify this relation further, we note that the average mass density of hydrogen in the Sun is roughly $1$ g/cm$^3$. 
In other words, $N_H\simeq 6\times 10^{53}$ given the volume of the Sun approximately at $10^{33}$ cm$^3$.    
Using this value of $N_H$ and the relation  $N_{\chi}(\tau^{\rm eq}_{\chi})=\sqrt{V_{\odot}C_c/\sigma_{\chi\chi}\bar{v}}$ with $C_c\simeq 5.4\times 10^{65}(\sigma^{\rm SI}_{\chi p}/{\rm cm}^2){\rm s}^{-1}$ and $\bar{v}\simeq 900$ km/s, we obtain $r\simeq 10^9\sqrt{\sigma^{\rm SI}_{\chi p}/\sigma_{\chi\chi}}$. We are interested in the parameter 
range that gives $r<1$, i.e., $\sigma^{\rm SI}_{\chi p}/\sigma_{\chi\chi}< 10^{-18}$. In this case, 
the trapped DMs reach to the thermal equilibrium among themselves before the energy exchange between the trapped DMs and the surrounding nuclei
becoming efficient. Besides the condition $r<1$, we need to ensure the ratio $C_s^2/4C_cC_a\equiv 1.9\times 10^3 (\sigma_{\chi \chi}/\sigma^{\rm SI}_{\chi p})(\sigma_{\chi\chi}/{\rm cm}^2)$ is much greater than unity for the consistency of our argument. The cross section combination $(\sigma^{\rm SI}_{\chi p},\sigma_{\chi\chi})=(10^{-45} {\rm cm}^2, 10^{-23} {\rm cm}^2)$
is an example of satisfying both $r<1$ and $C_s^2/4C_cC_a\gg 1$. With these parameters, we have $\tau^{\rm eq}_{\chi}=4.5\times 10^{13}$ s, which is much shorter than 
the age of the Sun,  $\tau_{\odot}\approx 10^{17}$ s. For spin-dependent cross section, we found 
$r\simeq 1.8\times 10^8\sqrt{\sigma^{\rm SD}_{\chi p}/\sigma_{\chi\chi}}$ and  $C_s^2/4C_cC_a=10^5 \times (\sigma_{\chi \chi}/\sigma^{\rm SD}_{\chi p})(\sigma_{\chi\chi}/{\rm cm}^2)$.  Since $C_c\simeq 10^{64}\times (\sigma^{\rm SD}_{\chi p}/{\rm cm}^2){\rm s}^{-1}$, the slightly different cross section combination  $(\sigma^{\rm SD}_{\chi p},\sigma_{\chi\chi})=(5\times 10^{-44} {\rm cm}^2, 10^{-23} {\rm cm}^2)$ gives the same $\tau^{\rm eq}_{\chi}$ as the previous spin-independent case.

The above discussions justify the thermal equilibrium state of DM in the early stage of capture for 
certain combinations of $\sigma_{\chi p}$\footnote{In the subsequent discussions we shall only focus on 
spin-independent cross section since it is better constrained by direct detection experiments. We drop the superscript SI henceforth for simplicity. } and $\sigma_{\chi\chi}$.  
We can write $E_{\chi}(t)=sk_{B}T_{\chi}(t)/2$ in Eq.~(\ref{thermaltransport}) 
where $s$ is the degree of freedom of each DM. 
We recapitulate the meaning of various coefficients in Eq.~(\ref{thermaltransport}). The coefficient $J_{c}$ describes the energy flow due to DM capture caused by DM-nucleus scattering,
$J_\chi$ describes the energy exchange between DMs and
nuclei, $J_{s}$ is related to the energy flow due to DM capture caused by DM self-interaction, and $J_{a}$ is related to 
the energy dissipation due to DM annihilation. All the quantities are positive except $J_\chi$ which depends on the difference between the solar core temperature $T_{c}$ and the DM temperature $T_{\chi}$.
In particular, $J_\chi < 0$ when $T_\chi > T_c$. 

When the Sun sweeps across the MW halo, there are
collisions between DM and the solar nuclei. If the DM velocity is smaller than the solar escape velocity after the collision, the DM will be gravitationally bounded within the Sun. The velocity of an infalling DM at the shell with radius $r$ inside the Sun 
is $w =\sqrt{u^{2}+v_{\rm esc}^{2}(r)}$. Here $u$ is the DM velocity in the halo and
$v_{\rm esc}^{2}(r)$ is the solar escape velocity at such shell. 
The kinetic energy of the trapped DM with an average over the energy loss  in the capture process is given by
\begin{equation}\label{ebar}
\bar{E}=\frac{m_{\chi}}{4}\left(\frac{m_{\chi}-m_{\rm A}}{m_{\chi} + m_{\rm A}}\right)^{2}  u^{2}   + \frac{m_{\chi}}{2}\frac{(m_{\chi}^{2} + m_{\rm A}^{2})}{(m_{\chi} + m_{\rm A})^{2}} v_{{\rm esc}}^{2}(r), 
\end{equation}
where $m_{\chi}$ is the DM mass and $m_{\rm A}$ is the mass of nuclei A. We have $\bar{E}=m_{\chi}v_{{\rm esc}}^{2}(r)/4$ if it is due to DM-DM scattering, i.e., $m_{\rm A} = m_{\chi}$. We note that the collisions among DMs redistribute the DM kinetic energies such that 
the average DM kinetic energy becomes $E_{\chi}(t)$ in Eq.~(\ref{thermaltransport}) when the thermal equilibrium is reached. The relation between $\bar{E}$ and $E_{\chi}(t)$ will be discussed later.
The energy flow per shell volume is
\begin{align}
 \frac{dJ_{c}}{dV}= & \int n_{\rm A}\sigma_{\chi\rm A}v_{{\rm esc}}^{2}(r)\frac{f(u)}{u}\nonumber \\
&\times\left[1-\frac{(m_{\chi} - m_{\rm A})^{2}}{4m_{\chi}m_{\rm A}}\frac{u^{2}}{v_{{\rm esc}}^{2}(r)}\right]\bar{E}du~,\label{eq:dLc/dV}
\end{align}
where $n_{\rm A}$ and $\sigma_{\chi \rm A}$ are the nuclei number density and the DM-nuclei
cross section, respectively. The DM velocity in the halo is assumed to follow Maxwell-Boltzmann distribution, i.e.,
\begin{eqnarray}
f(x) = \sqrt{\frac{6}{\pi}}\frac{\rho_{o}}{m_{\chi}\bar{v}}x^2e^{-x^2}e^{-\eta^2}\frac{\sinh(2x\eta)}{x\eta}~,
\end{eqnarray}
where $x^2 = 3(u/\bar{v})^2/2$ and $\eta^2 = 3(v_{\odot}/\bar{v})^2/2$ with $\bar{v} \approx 270~{\rm kms^{-1}}$ the DM dispersion velocity in the halo and $v_{\odot} = 220~{\rm kms^{-1}}$ the relative velocity between the Sun and the MW. The density $\rho_{0} $ is the DM local density\footnote{A slightly larger local DM density was derived in Ref.~\cite{Salucci:2010qr}.} taken to be $0.3~{\rm GeV/cm^3}$.       
The DM capture by Sun's nuclei gives 
\begin{equation}
J_{c}=\xi\sum_{\rm A}b_{\rm A}\frac{(m_{\chi}^{2}+m_{\rm A}^{2})}{(m_{\chi}+m_{\rm A})^{2}}\left(\frac{\sigma_{\chi \rm A}}{{\rm pb}}\right)\left\langle \phi_{\rm A}^{2}\right\rangle, \label{eq:J_c_Sun}
\end{equation}
where $b_{\rm A}$ is the number fraction of nuclei $\rm A$ in the Sun and $\left\langle \phi_{\rm A}^{2}\right\rangle$ is the average gravitational potential square contributed by nuclei A. The quantity $\xi$ is given by
\begin{align}
\xi & \equiv\sqrt{\frac{3}{8}}N_{\odot}\rho_{0}\frac{v_{{\rm esc}}(R_{\odot})}{\bar{v}}v_{{\rm esc}}^{3}(R_{\odot})\frac{{\rm erf}(\eta)}{\eta}\nonumber \\
 & \approx1.2\times10^{23}\,{\rm GeV}\,{\rm s}^{-1}\,\left(\frac{\rho_{0}}{0.3\,{\rm GeV}/{\rm cm}^{3}}\right)\left(\frac{270\,{\rm km}/{\rm s}}{\bar{v}}\right).
\end{align}
Similarly, the energy flow due to self-capture $J_{s}$ can be derived by making the replacements $m_{\rm A} \rightarrow m_{\chi}$ and $n_{\rm A} \rightarrow n_{\chi}$ ($n_{\chi}$ is the DM number density in the Sun). 
We obtain
\begin{equation}
J_{s} \approx \sqrt{\frac{3}{32}}\rho_{0}\sigma_{\chi\chi}\frac{{\rm erf}(\eta)}{\eta}\frac{v_{{\rm esc}}(R_{\odot})}{\bar{v}}v_{{\rm esc}}^{3}(R_{\odot})\left\langle \phi_{\chi}\right\rangle ^{2}.
\end{equation}
Since DMs are populated within a spherical region with a radius $R\simeq 0.1 R_{\odot}$, we can take
$\langle \phi^2_{\chi} \rangle \approx \langle \phi_{\chi} \rangle^2$ where  $\left\langle \phi_{\chi}\right\rangle =5.1$ is the average gravitational potential of the DM.  

The energy of captured DM could be dissipated due to annihilation. 
The energy flow due to this process is  
\begin{equation}
J_{a}=\frac{\int4\pi r^{2}n_{\chi}^{2}(r)E_{\chi}(t)dr}{(\int4\pi r^{2}n_{\chi}(r)dr)^{2}}\left\langle \sigma v\right\rangle 
\end{equation}
with $\left\langle \sigma v\right\rangle $ the thermally-averaged DM annihilation cross section.  
Numerically, we have
\begin{equation}
J_{a}(t)\approx 7.5\times10^{-65}\,{\rm GeV}\,{\rm s}^{-1}\left(\frac{sm_{\chi}}{10\,{\rm GeV}}\right)^{3/2}\left(\frac{E_{\chi}(t)}{{\rm GeV}}\right)^{-1/2}, \label{eq:E_a}
\end{equation}
where $s$ is the degree of freedom of each DM. We shall take $s=3$ for the numerical analysis later. 

Finally, the captured DMs continuously exchange energies with the solar nuclei. The energy transport due to collisions between
single DM and the surrounding nuclei in the Sun is given by~\cite{Spergel:1984re}
\begin{eqnarray}\label{Jcontact}
J_{\chi}&=&8\sqrt{\frac{2}{\pi}}\rho_{c}m_{\chi}\frac{k_B(T_{\odot}-T_{\chi})}{(m_{\chi}+m_{\rm A})^{2}} \nonumber \\ 
&\times&\sum_{\rm A}f_{\rm A}\sigma_{\chi \rm A}\left(\frac{m_{\rm A}k_BT_{\chi}+m_{\chi}k_BT_{\odot}}{m_{\chi}m_{\rm A}}\right)^{1/2}, \nonumber \\
\end{eqnarray}
where $\rho_{c} \approx 110~{\rm g/cm^{3}}$ is the core density of the Sun, $f_{{\rm A}}$ is the mass fraction of nuclei A. We note that the above relation is derived by taking $s=3$. 
\begin{figure*}[t]
\begin{centering}
\includegraphics[width=0.33\textwidth]{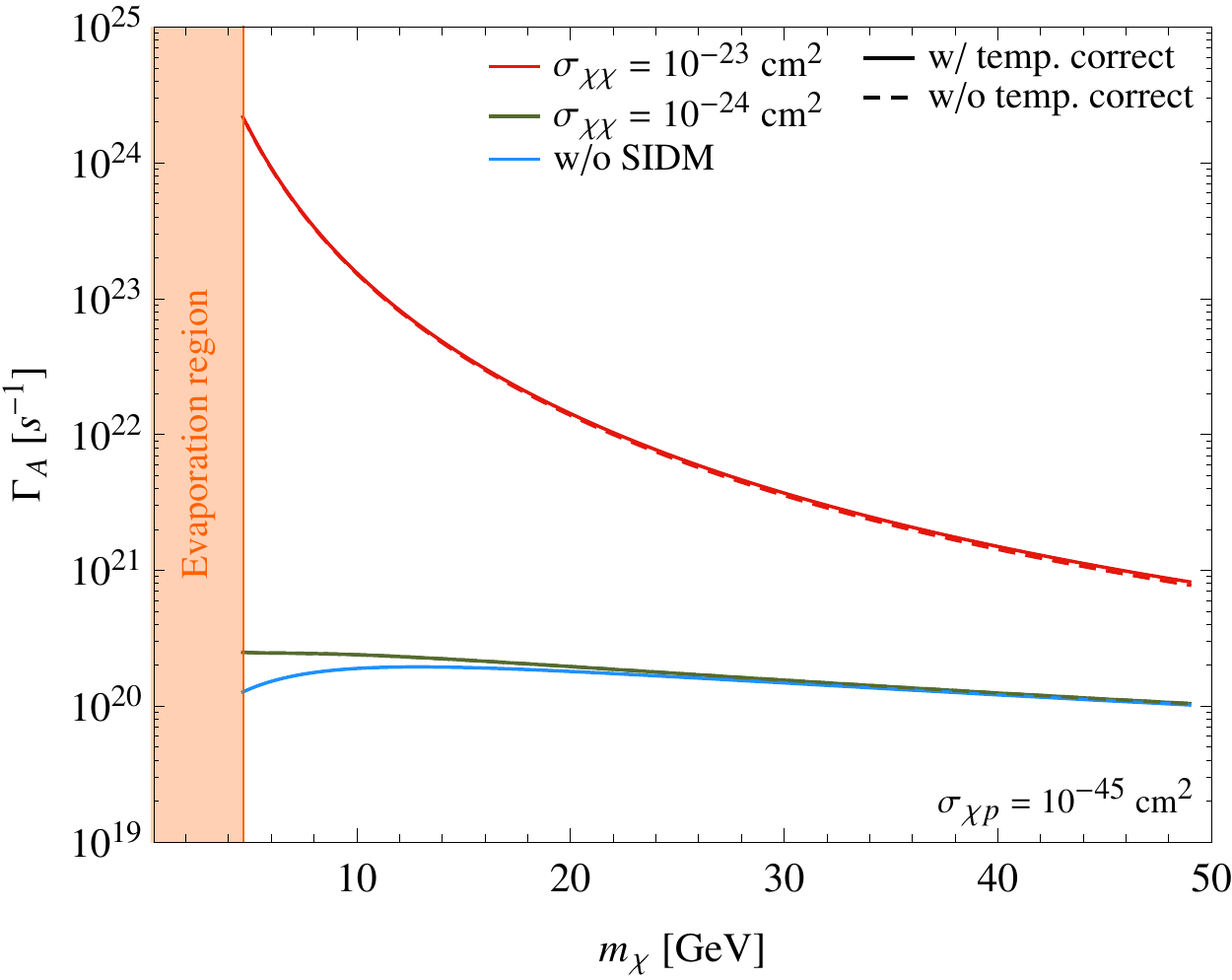}\quad\includegraphics[width=0.33\textwidth]{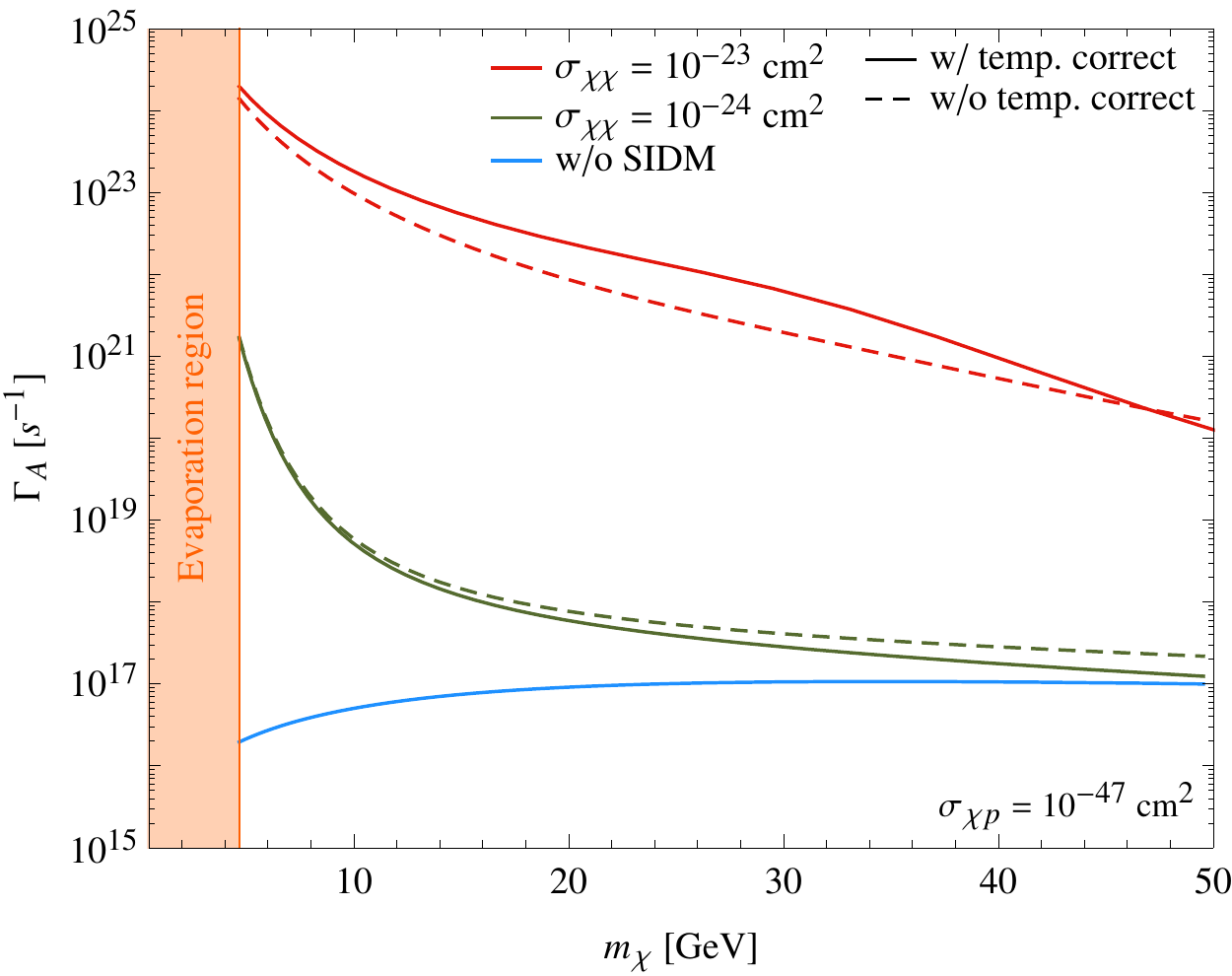}
\par\end{centering}
\protect\caption{Total annihilation rate $\Gamma_A$ for $\sigma_{\chi p} = 10^{-45}~{\rm cm^2}~\rm{and}~10^{-47}~{\rm cm^2}$ with $\sigma_{\chi\chi}= 10^{-23}~{\rm cm^2}$ and $10^{-24}~{\rm cm^2}$. 
Dashed lines are those results calculated with $T_{\chi}=T_c$. \label{annihilation}}
\end{figure*}
We have numerically solved Eqs.~(\ref{eq:N_evol}) and (\ref{thermaltransport}). 
Both equations run from $t_0=10^{13}$ s, which is roughly the time scale for DMs in the Sun to thermalize according to our earlier arguments. As a first approximation, the initial condition for $E_{\chi}(t)$ at $t_0=10^{13}$ s is taken to be $\bar{E}$ in Eq.~(\ref{ebar}) 
with $u^2$ and $r$ averaged. We point out that $\bar{E}$ is the DM average kinetic energy before the thermalization. When 
DMs reach to the thermal equilibrium, they are populated more closely to the solar core. Hence one expects $E_{\chi}(t_0)> \bar{E}$ by the conservation of mechanical energy, since the gravitational potential energy of
the thermalized DM is in general smaller than that of DM before thermalization. Nevertheless $E_{\chi}(t_0)$ and $\bar{E}$ differ within a factor of two, 
and in fact the DM temperature in the current epoch is independent 
of $E_{\chi}(t_0)$ as we shall argue later.

The evolutions of DM number and DM temperature are presented in Fig.~\ref{temperature}. 
We take $\sigma_{\chi p}=10^{-45},~10^{-47}~{\rm cm^2}$ and $\sigma_{\chi \chi}=10^{-23},~10^{-24}~{\rm cm^2}$ for $m_{\chi}= 5,~10,~50$ GeV, respectively as the benchmark points.
It is clearly seen that $N_{\chi}$ at $t\simeq 10^{13}$ s is indeed far below the maximal $N_{\chi}$. This justifies our assumption earlier for deriving $\tau_{\chi}^{\rm eq}$. 
We also see that the DM accumulation is enhanced by DM self interactions.
In particular, the $N_\chi$ enhancement is more significant for smaller $m_{\chi}$
and it is very sensitive to the ratio $\sigma_{\chi \chi}/\sigma_{\chi p}$.

Fig.~\ref{temperature} shows that DM temperature is higher than the temperature of solar core in the early stage 
for different DM masses. 	
The DM temperature $T_\chi$ drops with time due to energy flow from DMs to solar nuclei. However, $T_\chi$ is not always approaching to $T_c$. Those cases with
$\sigma_{\chi p} =10^{-47}~{\rm cm^2}$ show such a behavior.
Particularly, $T_\chi/T_c \sim 3.5$ for $\sigma_{\chi\chi}=10^{-23}~{\rm cm^2}$ and
$m_\chi=50$ GeV in the  current epoch. We note that $J_aN_{\chi}^2$ caused by the DM annihilation does not affect the DM temperature. Hence DM temperature evolution is controlled by
parameters $J_c$, $C_c$, $J_s$, $C_s$ and $J_{\chi}$. However, as $N_{\chi}$ accumulates, the effect
by $J_c$ and $C_c$ become negligible. We thus have
\begin{equation}
\frac{dE_{\chi}(t)}{dt} \approx J_\chi+J_{s}-C_sE_{\chi}(t). \label{DM_temp}
\end{equation} 
We observe that $E_{\chi}(t)$ or $T_{\chi}(t)$ approaches to a constant value when the right hand side of Eq.~(\ref{DM_temp}) vanishes. This implies $J_{\chi}\to 0$, i.e., $T_{\chi}\to T_c$, if the self-interaction induced coefficients $J_s$ and $C_s$ are
negligible. On the other hand, $T_{\chi}$ does not approach to $T_c$ if DM self-interaction cannot be neglected.   
We note that the gap between $T_{\chi}$ and $T_c$ increases for heavier DM. This is because $J_{\chi}$ scales as $1/m_{\chi}$for heavy DM. Thus, to balance $J_s-C_sE_{\chi}(t)$, $T_{\chi}-T_c$ must be enhanced to compensate the $1/m_{\chi}$ suppression in $J_{\chi}$. We argue that the DM temperature $T_{\chi}$ in the current epoch is fixed  once the values for $\sigma_{\chi\chi}$, $\sigma_{\chi p}$ and $m_{\chi}$ are given. Essentially, $T_{\chi}(t_{\odot})$ is the solution to $J_\chi+J_{s}-C_sE_{\chi}(t_{\odot})=0$ with $J_{\chi}$ the function of  
 $\sigma_{\chi p}$, $m_{\chi}$ and $E_{\chi}(t)$, and both $J_s$ and $C_s$ the functions of $\sigma_{\chi\chi}$. Clearly the initial condition $E_{\chi}(t_0)$
 does not affect $T_{\chi}(t_{\odot})$.  Indeed we have varied  $E_{\chi}(t_0)$ within a reasonable range in our numerical analysis and found that $E_{\chi}(t)$ ($T_{\chi}(t)$) approaches to the same fixed point.  

We remark that the thermal equilibrium time scale $\tau^{\rm eq}_{\chi}$, which we choose as $t_0$, is obtained by a rough estimation. To study the effect 
from such an uncertainty, we also solve Eqs.~(\ref{eq:N_evol}) and (\ref{thermaltransport}) with $t_0=10^{14}$ s. We find that 
$T_{\chi}(t_{\odot})$ remains the same.    

It is important to note that $N_{\chi}$ depends on the DM temperature $T_{\chi}$. This is due to 
the dependence of $N_{\chi}$ on $C_a$. Hence the indirect detection signal could also be sensitive to 
$T_{\chi}$.   
We compare the annihilate rate $\Gamma_A\equiv C_a N_\chi^2 /2$ calculated with precise $T_{\chi}$ with that
computed with the assumption $T_{\chi} = T_c$. The result is presented in Fig.~\ref{annihilation}.
DM temperature affects both $N_{\chi}$ and $C_a$ in $\Gamma_A$.  
With $T_{\chi}> T_c$, we have $C_a(T_\chi)< C_a(T_c)$ because $C_a \propto T_\chi^{-3/2}$ \cite{Griest:1986yu}. As a result, we can easily see that $N_\chi(T_{\chi})>N_\chi(T_c) $ from Eq.~(\ref{eq:N_evol}).
As long as the enhancement on $N_\chi^2$ can 
overwhelm the suppression on $C_a$, the annihilation rate  $\Gamma_A$ would be enhanced by adopting the precise 
DM temperature $T_{\chi}$. In the limit $C_s^2\gg 4C_cC_a$, $\Gamma_A\to C_s^2/2C_a$ when $N_{\chi}$
in the Sun reaches to the maximum. One has $\Gamma_A(T_{\chi})> \Gamma_A(T_{c})$ as a result. This is 
seen from the right panel of Fig.~\ref{annihilation} with $\sigma_{\chi p} =10^{-47}~{\rm cm^2}$ and $\sigma_{\chi\chi}=10^{-23}~{\rm cm^2}$. However, with $\sigma_{\chi\chi}=10^{-24}~{\rm cm^2}$
and $\sigma_{\chi p}$ kept the same, $N_{\chi}$ is still growing, i.e., $N_{\chi}\simeq C_c t_{\odot}$.
Hence $\Gamma_A\simeq C_aC_c^2t_{\odot}^2/2$, which implies $\Gamma_A(T_{\chi})< \Gamma_A(T_{c})$.
We presented such a scenario in the right panel of Fig.~\ref{annihilation} as well. 

We have taken the Sun as an example to illustrate the possible temperature difference between DMs and 
their surrounding medium. We stress that the derivation in this paper is generally applicable to other massive celestial objects.
The only required information is the DM dispersion velocity, the DM local density, and physical properties of the celestial object.          
In summary, we have derived and solved the thermal transport equation for DMs trapped in the Sun for the first time. 
In the SIDM scenario, the DM temperature and the core temperature of the Sun
could be different. We have shown that DM annihilation rate is sensitive to the DM temperature. 
Hence it is imperative to adopt the precise DM temperature to calculate the indirect DM event rates.

\section*{Acknowledgement} 
CSC is supported in part by National Center for Theoretical Sciences, Taiwan, R.O.C., and in part by the Ministry of Science and Technology (MOST), Taiwan, R.O.C. under the Grant No.~104-2112-M-001-042-MY3; GLL and YHL are supported by MOST, Taiwan under 
Grant No.~103-2112-M-009-018.

\end{document}